\documentclass[paper]{ieice}
\usepackage{booktabs}
\usepackage{graphicx}
\usepackage{nicefrac}
\usepackage[T1]{fontenc}
\usepackage{soul}
\usepackage{xspace}
\usepackage{flushend}  
\usepackage{color}
\usepackage{url}
\usepackage{booktabs}
\usepackage[table,xcdraw]{xcolor}
\usepackage{multirow}
\usepackage{listliketab}
\usepackage{amssymb}
\usepackage{algorithmic}
\usepackage{booktabs}
\usepackage{comment}
\usepackage{amsmath}
\usepackage{pgfplots}
\usepackage{bbding}
\usepackage{listings}
\usepackage{float}
\usepackage{tabularx}
\usepackage{cite}
\usepackage{rotating}
\usepackage[table,xcdraw]{xcolor}
\usepackage[square,sort,comma,numbers]{natbib}
\usepackage{tcolorbox}
\usepackage{ulem}
\usepackage{censor}
\usepackage{tikz}
\usepackage{amssymb} 
\usepackage{amsfonts}
\usepackage{graphicx,subfigure}
\usepackage{xparse}
\usepackage[colorinlistoftodos]{todonotes}
\usepackage{ulem}
\usepackage{amssymb}
\usepackage{lmodern}
\usepackage{mathtools, nccmath}
\usepackage{venndiagram}
\usepackage[symbol]{footmisc}

\usepackage{balance}
\usepackage{esvect}

\usepackage{xurl}
            
\usepackage[linesnumbered,ruled,vlined]{algorithm2e}

\SetCommentSty{mycommfont}

\SetKwInput{KwInput}{Input}                
\SetKwInput{KwOutput}{Output}              

\makeatletter
\def\@xfootnote[#1]{%
  \protected@xdef\@thefnmark{#1}%
  \@footnotemark\@footnotetext}
\makeatother

\setcitestyle{square}

\newcommand{\Maven}{Maven}
\newcommand{\npm}{npm}
\newcommand{\NuGet}{NuGet}

\newcommand{\RqFour}{RQ$_1$: \textbf{PM Issues Faced by End-users}- \textit{What types of PM issues do end-users face? }}
\newcommand{\RqFive}{RQ$_2$: \textbf{Underlying Causes of PM Issues}-\textit{What are the underlying causes of PM issues? }}
\newcommand{\RqSeven}{RQ$_3$: \textbf{Information Need to Resolve PM Issues}-\textit{What information is needed to resolve PM issues? }}

\usepackage{tcolorbox}
\usepackage{booktabs}
\usepackage{pgfplots}
\pgfplotsset{compat=newest}
\usepackage{multirow} 
\usepackage{pdflscape}
\usepackage{tikz}
\usepackage{progressbar}

\newcommand\syful[1]{{\textcolor{black}{#1}}}

\usepackage[colorinlistoftodos]{todonotes}

\field{}
\vol{}
\no{}
\SpecialSection{Empirical Software Engineering}

\title{An Empirical Study of Package Management Issues \\ via Stack Overflow}

\authorlist{%
\authorentry{Syful Islam}{n}{nstu}\MembershipNumber{}
\authorentry{Raula Gaikovina Kula}{n}{naist}\MembershipNumber{}
\authorentry{Christoph Treude}{n}{UM}\MembershipNumber{}
\authorentry{Bodin Chinthanet}{n}{naist}\MembershipNumber{}
\authorentry{Takashi Ishio}{m}{naist}\MembershipNumber{1301152}
\authorentry{Kenichi MATSUMOTO}{f}{naist}\MembershipNumber{8925358}
}

\affiliate[nstu]{The author is with the 
Noakhali Science and Technology University, Bangladesh.}

\affiliate[naist]{The authors are with the Graduate School of Science and Technology, Nara Institute of Science and Technology, Ikoma-shi, 630-0192, Japan.
}

\affiliate[UM]{The author is with the 
University of Melbourne, Australia.}

\received{2015}{1}{1}
\revised{2015}{1}{1}

\begin{document}
\maketitle
\begin{summary}
\noindent The  package manager (PM) is crucial to most technology stacks, acting as a broker to ensure that a verified dependency package is correctly installed, configured, or removed from an application.
Diversity in technology stacks has led to dozens of PMs with various features.
While our recent study indicates that package management features of PM are related to end-user experiences, it is unclear what those issues are and what information is required to resolve them. 
In this paper, we have investigated PM issues faced by end-users through an empirical study of content on Stack Overflow (SO).
We carried out a qualitative analysis of 1,131 questions and their accepted answer posts for three popular PMs (i.e., \Maven, \npm, and \NuGet\ ) to identify issue types, underlying causes, and their resolutions.
Our results confirm that end-users struggle with PM tool usage (approximately 64-72\%).
\syful{We observe that most issues are raised by end-users due to lack of instructions and errors messages from PM tools. In terms of issue resolution, we find that external link sharing is the most
common practice to resolve PM issues. Additionally, we observe that links pointing to useful resources (i.e., official documentation websites, tutorials, etc.) are most frequently shared, indicating the potential for tool support and the ability to provide relevant information for PM end-users.}

\end{summary}
\begin{keywords}
Package Manager, End-user Issues,  Stack Overflow
\end{keywords}

\section{Introduction}
\syful{Package manager (PM) is crucial to most the technology stacks in software development, especially when building a web or mobile application.
Using a package as the third-party dependency in software applications is prominent, with more than 5 million open source packages available via PMs~\cite{Web:libraries}.
In 2020, GitHub showed its support for third-party package usage when it acquired the Node.js PM (i.e., {\npm}), which serves over 1.3 million packages to roughly 12 million end-users, and is constantly growing each day~\cite{Web:npmStat}.}

\begin{figure*}[t]
    \centerline{\includegraphics[width=0.9\linewidth]{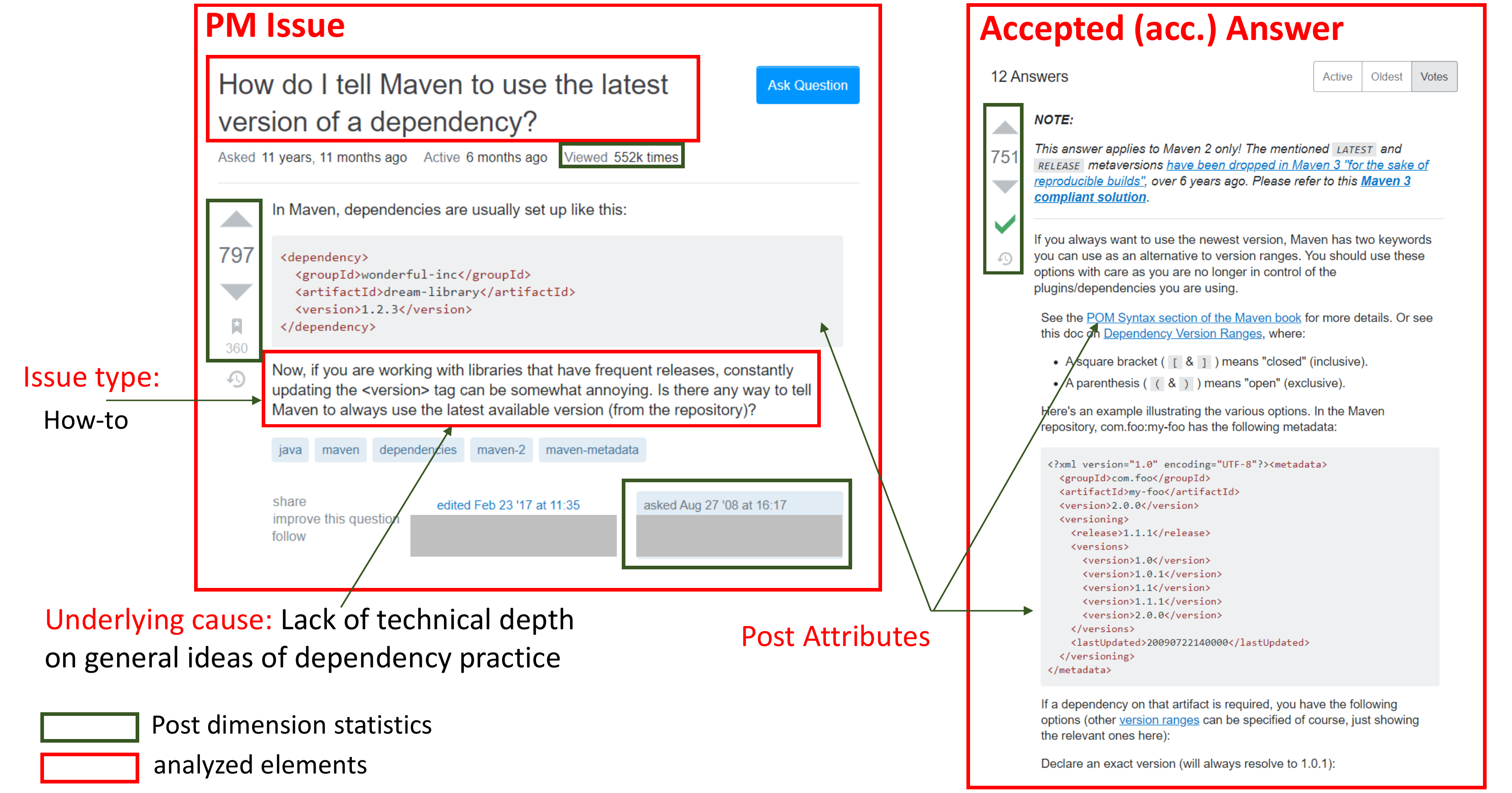}}
    \caption{\syful{A PM-related issue that provides an exemplary overview of the elements analyzed in our study. RQ$_1$ and RQ$_2$ analyze the issue type and underlying cause from the PM-related post text. Finally, RQ$_3$ analyze the relationship between the PM question and the accepted answer posts to identify the information need for issue resolution.}}
    \label{fig:example}
\end{figure*}

\syful{A PM acts as an intermediary broker between an application and a third-party package dependency, to ensure that a verified package is correctly installed, configured, or removed from an application.
PMs were developed as an automated solution to deal with version compatibility and build issues that arise when an application adopts numerous dependencies~\cite{dietrich2019dependency}.}
Diversity in technology stacks has led to a variety of PMs with different features. 
For instance, npm brokers packages that run in the node.js based environment and are written in JavaScript.
Recent studies have mainly investigated dependency management and migration \cite{Cox-ICSE2015,bogart2016break,kula2018developers,decan2018impact,decan2019empirical}.
In our study~\cite{islam2021contrasting}, we explored thirteen PMs to understand whether their package management features correlate with the experience of their end-users.  While our recent study indicates that package management features of PM are related to end-user experiences, it is unclear what those issues are and what information is required to resolve them. 

\syful{Building on our previous study, we empirically investigate the issues that end-users face when using PMs and the information needed to resolve them.
Since it is common practice for end-users to use Stack Overflow to post questions about issues faced during software development, we leverage this data source for our work.
First we conduct an in-depth qualitative analysis on 1,131 sample question posts on three PMs, i.e., {\Maven}, {\npm}, and {\NuGet} to explore PM issues in term of issue types and their underlying causes. Next, we investigate the information need to resolve PM issues through a mixed method analysis (i.e., both qualitative and quantitative).} Specifically, we investigated the following three research questions:
    \begin{itemize}
            \item \RqFour Our motivation is to gain a better understanding of the challenges that end-users face through an investigation of  PM issues. Using the question coding schema of Treude et al.~\cite{treude2011programmers}, we are able to classify each PM-related issue. \syful{As shown in Figure~\ref{fig:example}, the PM question can be classified as a how-to type issue in which an end-user asks for instructions. } 
           
            \item \RqFive Our motivation is to discover the underlying cause of end-users challenges when they use PM. By reading each PM-related issue, we are able to identify and categorize common underlying causes which explain why end-users struggle with using a PM. \syful{As shown in Figure~\ref{fig:example}, an example underlying cause of PM issue can be related to lack of technical depth on general ideas of dependency practice.}
            
            \item \RqSeven Our motivation is to identify useful information patterns that can assist in resolving PM issues. Using a mixed-method analysis between question and accepted answer posts, we are able to identify common patterns of information and determine whether they are related to a resolved PM issue. \syful{As shown in Figure~\ref{fig:example}, the build configuration file is identified as a PM-related post attribute which contributes to the issue resolution.} 
        \end{itemize}
        
   \syful{Our findings from RQ$_1$ show that, most PM issues are related to \texttt{how-to} questions followed by \texttt{error} messages, confirming end-users lack instructions and intuitive error messages when using PMs. These findings indicate that, end-users are interested in more specific guidance relating to intuitive instructions, concepts, and errors associated with PM. In addition, it is necessary to better understand what technical background knowledge end-users should have concerning PMs.  From RQ$_2$, we find that the underlying causes of
PM issues can be categorized into package management tool usage, general dependency practices, specific migration, and others. We observe that \texttt{package management tool
usage} is the most dominant underlying cause, confirming end-users report technical
issues on package management tool usage and not on
specific migrations of dependencies. These findings indicate the necessity of making it easier for end-users to find the information they need to resolve PM tool usage issues, for example, by providing good error messages or intuitive instructions. Finally, from RQ$_3$, we find that end-users post \texttt{external links} to resolve PM issues. We observed that links pointing to useful resources such as official documentations and tutorials are most frequently shared. This finding reveals the opportunities for tool support and strategies to find useful information for addressing PM issues. Researchers could develop approaches that suggest useful information links by automatically analyzing stack traces and logs to help speed up the issue-resolution process.}

The rest of the paper is organized as follows. Section~\ref{Methodology of Dataset Building} describes the methodology of dataset building for the target PMs. Experimental results and discussions are presented and discussed in Section~\ref{Results and Discussions}.  Section~\ref{implications} presents the implications of this work. Section~\ref{validity} presents the threats to validity of this work. Section~\ref{sec:related_work} presents the related works. Finally, section~\ref{conclusions} presents the conclusion of this work.

\section{Methodology of Dataset Building}
\label{Methodology of Dataset Building}
In this section, we explain the methodology of selecting PMs and their relevant question-and-answer post dataset building.

\noindent\textbf{Selecting PMs.} 
\syful{ For in-depth examination of package management issues faced by end-users, we focused primarily on PMs associated with web frameworks, owing to their increasing popularity for creating web applications, and digital systems. We selected the most popular PMs associated with the top eight web frameworks based on the Stack Overflow survey 2020\footnote{Stack Overflow Survey: \url{https://insights.stackoverflow.com/survey/2020}}. This resulted in three PMs: npm (i.e., jQuery, React.js, angular, express, Vue.js are web frameworks of JavaScript), NuGet (i.e., ASP.NET is web framework of C\# .NET) and Maven (i.e., Spring is web framework of Java).}

\noindent\textbf{Collecting PM Posts.}
\syful{Initially, we downloaded the Stack Overflow data dump version 2019-12-25 published on SOTorrent~\cite{DBLP:conf/msr/BaltesDT008}.}
To collect relevant PM posts, we utilized tag-based question post filtering which was also used by prior studies~\cite{abdellatif2020challenges, rosen2016mobile}. 
\syful{To filter out tags pertaining to the three PMs, first we collected question posts tagged with the keyword \texttt{package-managers}. This step resulted in 806 question posts. Then, we extracted all tags from the 806 question posts, resulting in 626 unique tags. We observed that, although some tags were rare (i.e., occurring only once) in the initial posts dataset, they had a strong relationship with the three PMs. The tag maven-2, for instance, appears only once in the initial PM post dataset as a coexisting tag, but the tag itself is associated with 5,568 posts related to maven. Hence, two authors manually filtered tags from 626 unique tags and identified 11 tags that are associated with three PMs (see Table~\ref{tab:tag_list}).
Finally, we applied these tags to collect PM posts.} \syful{Thus, we obtained PM question-and-answer dataset (i.e., D1) with 114,834  questions and 50,696 accepted answer posts. Afterward, we prepared representative sample dataset (i.e, D2), maintaining 95\% confidence level and a confidence interval of 5 for each PM. This resulted in a total of 1,131 sample question  from three PMs (i.e., \Maven:382, \npm:379, and \NuGet:370 posts) and their associated accepted answers. Table~\ref{tab:samples} provides an overview of the prepared datasets for characterization of PM issues, their underlying causes and the information need to resolve them. All data and scripts are made available through our replication package\footnote{Replication Package: \url{https://zenodo.org/record/7005818\#.Yv3H0OzP30o}}.}

\begin{table}[t]
\centering
 \caption{
 \syful{List of tags used to collect PM posts}
 }
    \label{tab:tag_list}
\begin{tabularx}{\linewidth}{p{0.9in}p{2.1in}@{}}
\hline
 Initial tag & Identified relevant tags  \\ \hline
 package-managers & npm, nuget, nuget-package, maven, npm-install, npm-scripts, npmignore, pnpm, npm-shrinkwrap, nuget-package-restore, maven-2  \\
 \hline
\end{tabularx}

\end{table}

\begin{table}[t]
\centering
\caption{
\syful{Summary of the PM posts dataset prepared for characterization of PM issues and information need to resolve them.}
}
 \label{tab:samples}
\begin{tabular}{@{}lrrlcc@{}}
\toprule
 & \multicolumn{2}{c}{Dataset (D1)} &  & \multicolumn{2}{c}{Sample Dataset (D2)} \\ \cmidrule(lr){2-3} \cmidrule(l){5-6} 
PM & \multicolumn{1}{c}{\#Question} & \multicolumn{1}{c}{\begin{tabular}[c]{@{}c@{}}\#Accepted\\ answer\end{tabular}} &  & \#Question & \begin{tabular}[c]{@{}c@{}}\#Accepted\\ answer\end{tabular} \\ \midrule
Maven & 74,657 & 33,710 &  & 382 & 173 \\
npm & 30,136 & 11,857 &  & 379 & 151 \\
NuGet & 10,041 & 5,129 &  & 370 & 182 \\ \midrule
Total & 114,834 & 50,696 &  & 1,131 & 506 \\ \bottomrule
\end{tabular}
\end{table}

\section{Results and Discussions}
\label{Results and Discussions}
\syful{In this section, we present the approaches used to analyze PM posts and the results obtained to answer our research questions.}

\subsection{PM Issues Faced by End-users ($RQ_1$)}
\label{subsection:Kind of Questions asked about Package Ecosystems}

\begin{figure*}[t]
    \centerline{\includegraphics[width=0.70\linewidth]{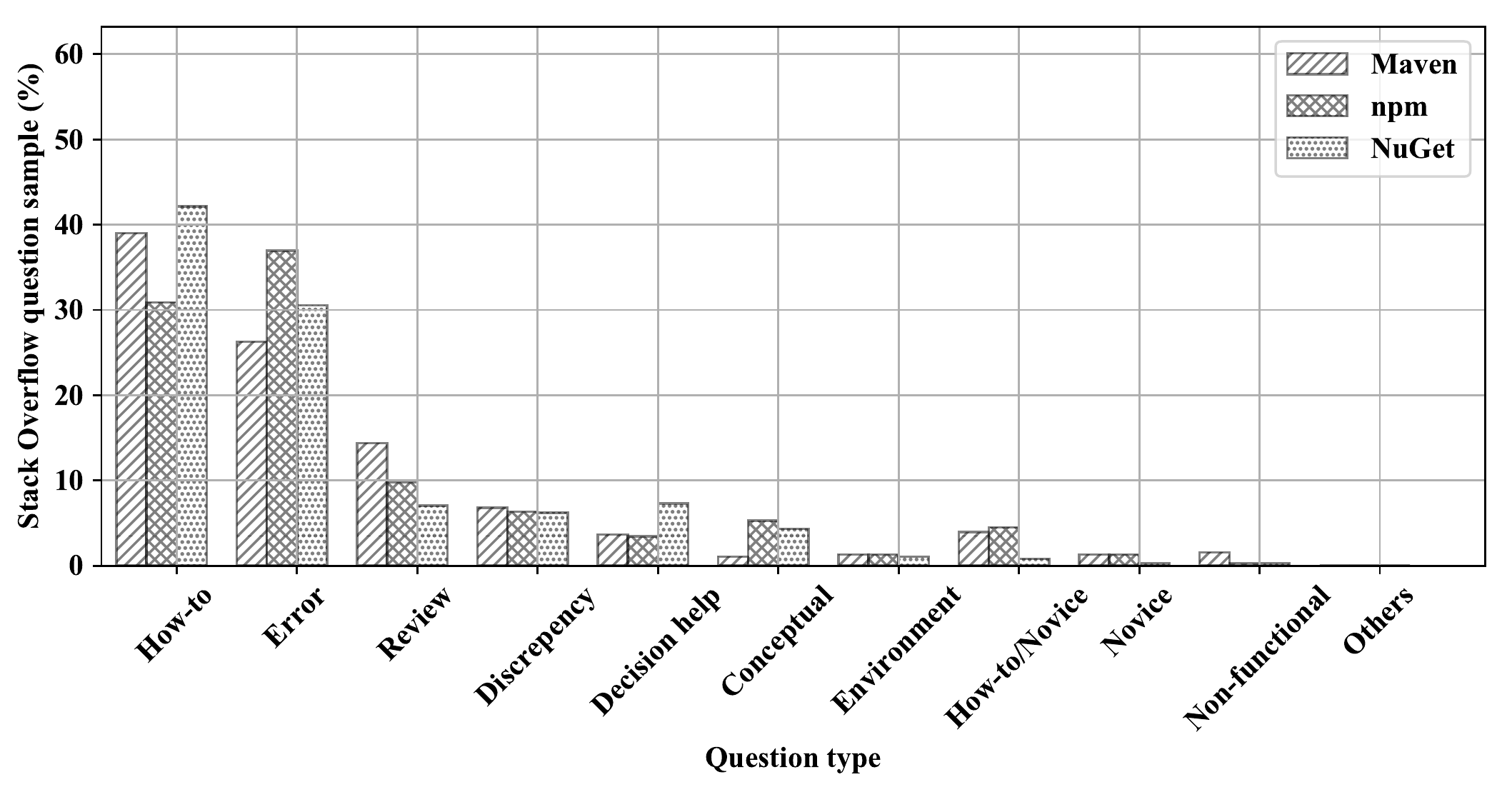}}
    \caption{Percentage of PM issues by question coding from Treude et al.~\cite{treude2011programmers}. Result shows that \texttt{`How-to'} and \texttt{`Error'} messages  are the most dominant issue faced by end-users.}
    \label{fig:RQ4}
\end{figure*}

\noindent\textbf{Approach. }
\syful{To answer RQ$_1$, we conducted a qualitative analysis of statistically representative sample questions included in dataset D2 (see Table~\ref{tab:samples}) for three PM tools, namely maven, npm, and NuGet. The question coding scheme proposed by Treude et al.~\cite{treude2011programmers} was adopted to annotate sample questions into PM issues faced by end-users.}
Details of the question coding scheme are described below:

\begin{itemize}
    \item \textit{How-to}: Posts that ask for instructions. \syful{For example: ``Title: How to get latest version number of an artifact and replace it in target file? (Id: 26223226)''. In this post, a developer asked for instruction about getting latest version an artifact.}
    \item \textit{Discrepancy}: Some unexpected behavior that the person asking about the PM post wants explained. \syful{For example: ``Title: Spring Boot JPA \& H2 Records Not Persisted (Id: 27843682)''. In this post, a developer faced some unexpted issues while running command \texttt{mvn spring-boot:run}.}
\item \textit{Environment}:  Posts about the environment either during development or after deployment. \syful{For example, ``NPM Windows Path Problems (Id: 25120982)''. In this post, a developer ask for solution of environment setting issue.}
\item \textit{Error}: Posts that include a specific error message. \syful{For example: ``Title: Error running Google App Engine quick start : POM for com.google.app engine:app engine-maven-plugin:jar:1.9.24 is missing (Id: 31576681)''. . In this post, a developer asked for solution of build failure caused after running command \texttt{mvn appengine:devserver}.}
\item \textit{Decision help}: Asking for an opinion. \syful{For example: ``Can I invoke a local bean into a ear file from a Javax-WS into a war file- apache-tomee-plus-1.7.4 (Id: 51072906)''. In this post, a developer asked for an opinion about invoking a local bean into a ear file from a Javax-WS into a war file-apache-tomee-plus.}
\item \textit{Conceptual}: Posts that are abstract and do not have a concrete use case. \syful{For example: ``Difference between mvn appengine:update and mvn appengine:deploy in Google App Engine (Id: 40094090)''. In this post, a developer asked for understanding the difference between mvn appengine:update and mvn appengine:deploy.}
\item \textit{Review}: Posts that are either implicitly or explicitly  asking for a review. \syful{For example: ``Is there any possibility of deleting libraries stored maven central? (Id: 25133985)''. In this post, a developer asked for a review about possibility of deleting libraries stored maven central.}
\item \textit{Non-functional}: Posts about non-functional requirements  such as performance or memory usage.  \syful{For example: ``Java project runs slow from JAR but fast from IDE (Id: 41861330)''. In this post, a developer asked for performance issue of Java project.}
\item \textit{Novice}: Often explicitly states that the person belong PM posts is a novice. \syful{For example: ``The missing package org.spring framework.web (Id: 12603723)''. In this post, a novice developer asked for a package missing issues while using Spring MVC.}
\item \textit{How-to/Novice}: Posts that belong to a novice asking for step by step tutorials. \syful{For example: ``Maven2 - POM configure issue in Windows (Id: 5087296)''. In this post, a novice developer asked for step by step instruction to configure maven POM file.}

\item \textit{Others}: Posts that don't fall in the above  categories.
\end{itemize}
\syful{In our annotation guidelines, we did not allow multiple categories for one question post.
 To ensure the quality of our classification, we performed a Kappa agreement check using 30 random samples among three authors.
Using the Kappa score calculator~\cite{viera2005understanding}, we checked the agreement level and find overall score 86.67\%, which was almost perfect. 
Confident with the agreement, two authors then continued to manually annotate the remaining sample posts.}

\noindent\textbf{Results. }
\syful{Figure~\ref{fig:RQ4} shows the result of our qualitative analysis. We observe that most PM issues are related to \texttt{How-to} questions followed by \texttt{Error} messages, confirming end-users lack instructions and intuitive error messages when using PMs.}
In {\Maven}, we find that \texttt{How-to} (39.01\%) question is most dominant, followed by \texttt{Error} (26.96\%) and \texttt{Review} (14.40\%). Similar trend is also shared by {\NuGet}, where \texttt{How-to} (42.16\%)  question is most dominant, followed by \texttt{Error} (30.54\%) and \texttt{Decision help} (7.30\%). On the other hand, in {\npm} we find that \texttt{Error} (36.94\%) message question is most dominant, followed by \texttt{How-to} (30.87\%) and \texttt{Review} (9.76\%). \syful{These findings indicate that, end-users are interested in specific guidance relating to intuitive instructions, concepts, and errors associated with PMs.}

\begin{tcolorbox}
\textbf{Answering RQ$_1$:}
The PM issues faced by end-users belong to question type how-to (i.e., around 31 to 42\% of question samples) and error messages (i.e., around 27 to 37\% of questions). End-users of Maven and  NuGet asked the most how-to type questions while npm end-users ask mostly error type questions.  
\end{tcolorbox}

\subsection{Underlying Causes of PM Issues ($RQ_2$)}
\label{underlying causes}

\noindent\textbf{Approach. }
\syful{To answer RQ$_2$, we conducted a qualitative analysis of statistically representative sample questions included in dataset D2 (see Table~\ref{tab:samples}) for three PM tools, namely maven, npm, and NuGet. An open coding strategy was adopted similar to the previous work by Hata et al.~\cite{HataICSE2019} in order to reveal the root causes of PM issues.} The open coding process consists of three steps.
 In step-1, the second author of this paper derived a draft list of underlying causes based on 30 randomly sampled posts. Then the first two authors used the draft list to label the sample questions collaboratively. During this step, the underlying causes were revised. The output of this step was 4 underlying causes including others. In step-2, the same two authors performed manual annotation of another 30 samples to make sure that no new causes appeared. The output of this step was the same 4 underlying causes.  In step-3, three authors discussed the coding results that were obtained in  step-2 and measured author agreement using Kappa score~\cite{viera2005understanding}.\syful{  We did not allow multiple categories for one question post under our annotation guidelines.
 After two rounds of manual annotation of 30 samples among three authors, we ended up with a Kappa score of 95.56\% (almost perfect).}
 \syful{Given reliability of the annotation guidelines, two authors manually annotated the remaining samples.} 
 The 4 underlying causes of PM issues are summarized below: 

 \begin{itemize}
     \item \textit{Specific migration}: \syful{Posts related to dependency updates like an upgrade or downgrade to a specific version of the package, moving to new environment (language/OS/etc), incompatibility of packages and other associated issues. \syful{For example, from the PM Post (Id: 50262939), we observe that an end-user struggle with issues to migrate his project written in Java8 to Java9.}}
 
\item \textit{Package management tool usage}: Posts related to technical details on package management systems such as installation, configuration of tools, and their associated costs. \syful{For example, from the PM Post (Id: 4811870), we observe that an end-user discusses about configuring the maven build tool to run multi-module project in a custom way.}

\item \textit{General dependency practices}: Posts related to lack of technical knowledge on general dependency management practice, bugs, efficiency, etc. For example, from the PM Post (Id: 30571), we observe that, end-users ask a question on the general ideas of maven packages dependency management.
  
\item \textit{Others}: Posts that are tagged with a PM but do not fall in the above three categories. For example, the PM post (Id: 38906885) does not fall in the above three categories.
 \end{itemize}

\begin{figure}[t]
    \centerline{\includegraphics[width=0.90\linewidth]{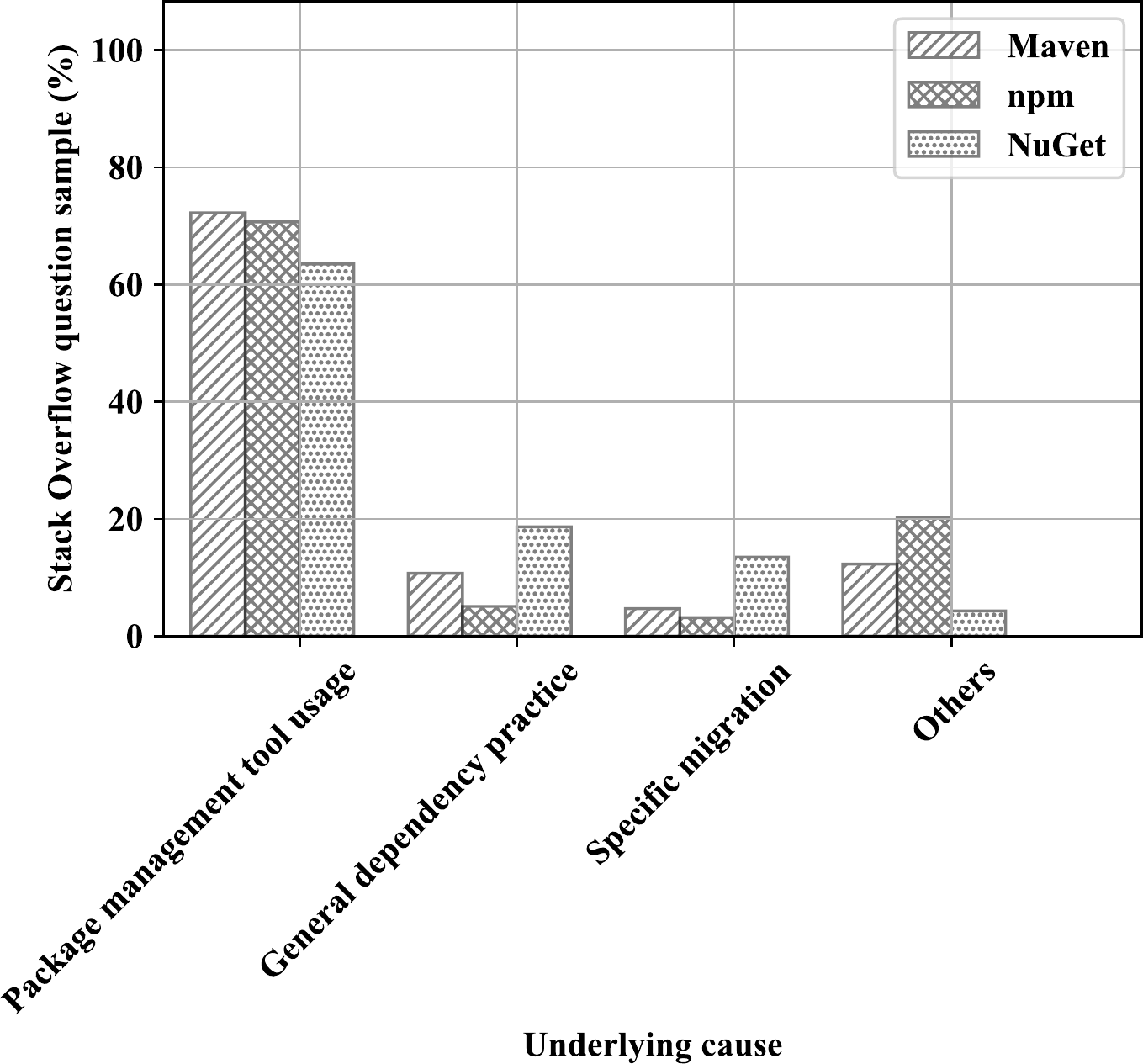}}
    \caption{Percentage of PM issues by their underlying cause. We find that \textit{`Package management tool usage'} is the most dominant underlying cause. }
    \label{fig:underlying_cuases}
\end{figure}

 \noindent\textbf{Results. }  
 \syful{Figure~\ref{fig:underlying_cuases} shows the result of our qualitative analysis.} We find that \texttt{Package management tool usage} is the most dominant underlying cause for all three PMs, confirming that end-users report technical issues on Package management tool usage and not on specific migrations of dependencies. The result is consistent with previous studies~\cite{nagaria2020software, mangul2019challenges, argelich2008cnf, decan2017empirical, abate2015mining}. They reported that the complexity of software tools~\cite{nagaria2020software}, installability of tools and packages~\cite{mangul2019challenges, decan2017empirical, abate2015mining, argelich2008cnf} are the root causes of the end-users struggle during application development and maintenance.
In detail, we find that \texttt{Package management tool usage} (72.25\%) is most dominant underlying cause for {\Maven}, followed by \texttt{General dependency practices} (10.73\%). \syful{The same trend is also evident in the other two PMs, namely, {\npm} and {\NuGet}.} In {\npm}, \texttt{Package management tool usage} (70.71\%) is most dominant underlying cause, followed by \texttt{General dependency practices} (5.80\%). In {\NuGet}, \texttt{Package management tool usage} (63.51\%) is most dominant underlying cause, followed by \texttt{General dependency practices} (18.65\%) and \texttt{Specific migration} (13.51\%). \syful{These findings indicate the necessity of making it easier for end-users to find the information they need to resolve PM tool usage issues, for example, by providing good error messages or intuitive instructions.}

\begin{tcolorbox}
    \textbf{Answering RQ$_2$}: 
   The underlying causes of PM issues can be categorized into package management tool usage, general dependency practices, specific migration, and others.
End-users ask questions that relate to package management tool usage (approximately 64-72\%), and not specific migration and general dependency practice.
\end{tcolorbox}

\subsection{Information Need to Resolve PM Issues ($RQ_3$)}

\noindent\textbf{Approach. }
\syful{To answer RQ$_3$, we conducted a mixed method analysis (i.e., both qualitative and quantitative) to explore useful information patterns between PM questions and their accepted answers. This process was completed
through two consecutive steps i.e., (i) Identify the most dominant attribute from PM posts, and (ii) Discover the useful information patterns from most dominant attribute. We have explained each step below.}

\begin{itemize}
    \item \syful{\textit{Step 1: Identify the most dominant attribute from PM posts. }  
    We conducted a qualitative analysis of statistically representative sample questions included in dataset D2 (see Table~\ref{tab:samples}) for three PM tools. An open coding strategy was adopted to identify the most dominant PM post attributes.
    The open coding process consists of three steps. Initially, the first author of this paper derived a draft list of attributes based on 30 randomly sampled posts. Then the two authors used the draft list to label the sample questions collaboratively. During this step, the attributes were revised. The output of this step was 6 attributes. In the next step, the same two authors performed manual analysis on another 30 samples to make sure that no new causes appeared. The output of this step was the same 6 attributes as the previous step. Finally, two authors discussed the coding results and measured authors agreement using Kappa score~\cite{viera2005understanding}. In our annotation guidelines, we allowed multiple attribute annotations for one question post. The overall kappa score among two authors for PM question posts attribute coding were source code 100\%, build configuration file 84.62\%, textual content 100\%, external link 93.33\%, cmd/log/output files 84.62\%, version information 76.92\%, respectively. 
Given reliability of the annotation guidelines, the first author annotated remaining sample posts. The 6 attributes of PM posts are summarized below:}
\begin{itemize}
    \item \textit{Source code}: The post includes code snippets as part of the information.
\item \textit{Build configuration file}: The post includes build configuration files as a part of information.
\item \textit{cmd/log/output file}: The post includes log files or output from the program.
\item \textit{External link}: The post includes external links as a part of information.  
\item \textit{Version information}: The post includes version information of a package or any software.
\item \textit{Textual content only}: The post includes only text as information. 
\end{itemize}

 \syful{After completing the annotation process, we analyzed the annotated samples to identify the most dominant attribute. The output of this step was the most dominant attribute  from the PM posts.}

\begin{figure*}[t]
\centering
     \subfigure[Questions with accepted answer]{
       \includegraphics[width=.40\linewidth]{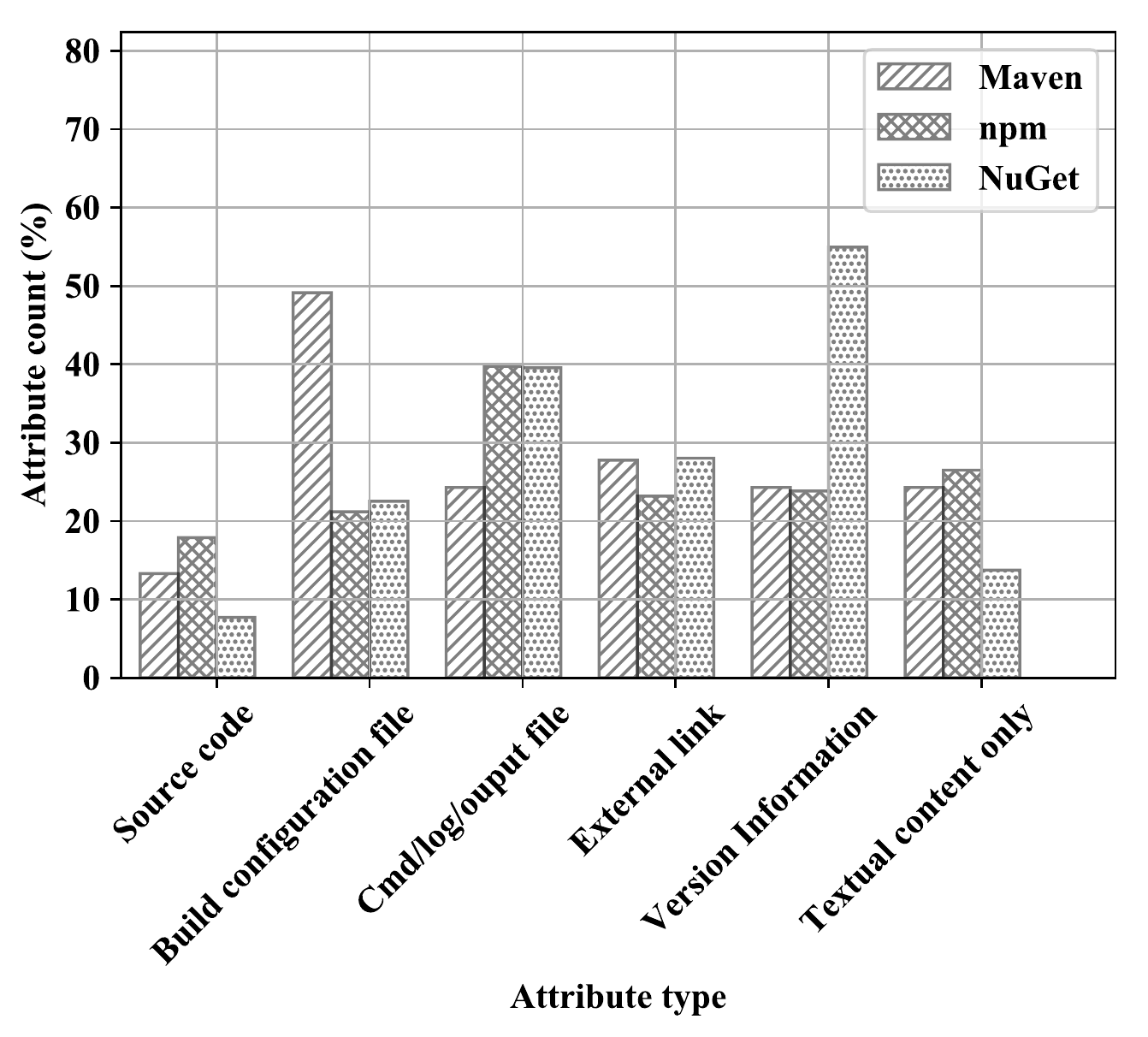}}
        \subfigure[Accepted answers]{
        \includegraphics[width=.40\linewidth]{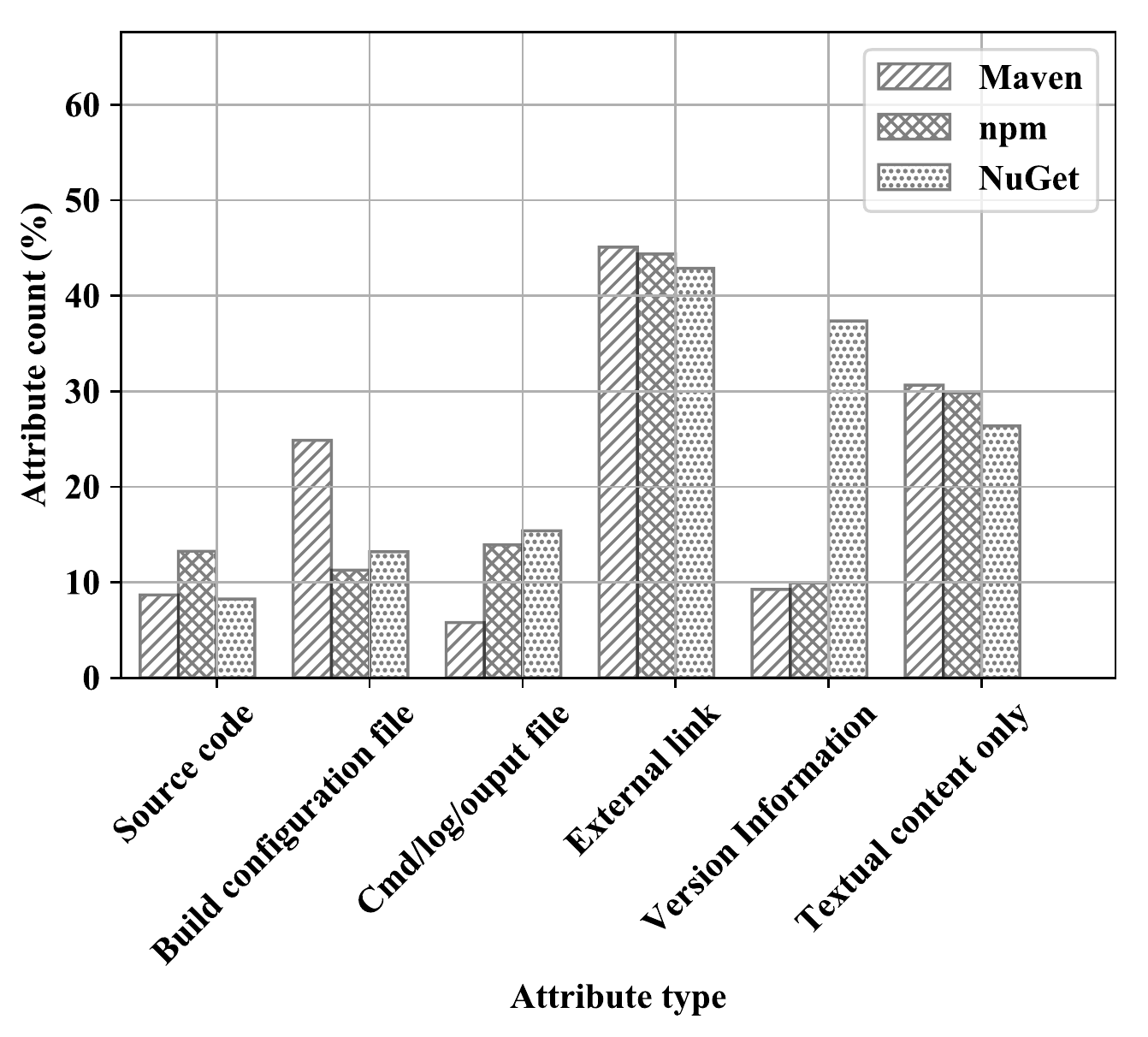}}
       \caption{Attribute analysis between PM question and their associated accepted answer posts. External link is the most dominant attribute in the accepted answer of PM issues.}
    \label{Fig:RQ6_attribute}
\end{figure*}

    \item \textit{Step2: Discover the useful information patterns from most dominant attribute. } \syful{We conducted a quantitative analysis on dataset D1 (see Table~\ref{tab:samples}) to discover useful information patterns for the most dominant attribute (i.e., output of step 1) between PM questions and their accepted answers. Here, we applied association rule mining on dataset D1 in order to increase the chance of discovering more useful information patterns. First, we extracted most dominant attribute information's from PM questions and their associated accepted answers by using regular expression. Next, we performed necessary pre-processing on the extracted information such as removing special characters, backslash etc.
Afterward, we applied the \textit{apriori} algorithm as implemented in the python package \texttt{mlxtend}\footnote{Mlxtend Package: \url{http://rasbt.github.io/mlxtend/api_subpackages/mlxtend.frequent_patterns/}}.
Finally, we filtered all rules with at least 5 data points and 45\% confidence. 
The output of this step was useful information patterns from most dominant attribute. }
\end{itemize}

\begin{table*}[t]
\centering
\caption{Aggregated association rules for the \texttt{external link} attribute extracted from PM questions (Q\_domain) and their accepted answers (A\_domain). We find that links refer to useful information's such as official documentations, tutorials, GitHub resources, and screenshots are most frequently shared by end-users to resolve PM issues.}
\label{tab:RQ6_rules}
\begin{tabular}{@{}llcc@{}}
\toprule
& Association rule & \multicolumn{1}{c}{Confidence} & \multicolumn{1}{c}{Support} \\ \midrule
 & Q\_maven.apache.org$\,\to\,$A\_maven.apache.org & 49.67\% & 300 \\
 & Q\_maven.apache.org, Q\_i.stack.imgur.com$\,\to\,$A\_maven.apache.org & 56.00\% & 14 \\
 & Q\_code.google.com, Q\_stackoverflow.com$\,\to\,$A\_github.com & 50.00\% & 9 \\
\Maven & Q\_docs.gradle.org$\,\to\,$A\_docs.gradle.org & 50.00\% & 7 \\
 & Q\_www.oracle.com$\,\to\,$A\_stackoverflow.com & 53.85\% & 7 \\
 & Q\_www.jfrog.com$\,\to\,$A\_www.jfrog.com & 58.33\% & 7 \\
 & Q\_code.google.com, Q\_github.com$\,\to\,$A\_github.com & 50.00\% & 6 \\ \midrule
 & Q\_github.com$\,\to\,$A\_github.com & 51.73\% & 359 \\
 & Q\_www.npmjs.com, Q\_github.com$\,\to\,$A\_github.com & 58.62\% & 34 \\
 & Q\_i.stack.imgur.com, Q\_github.com$\,\to\,$A\_github.com & 52.83\% & 28 \\
 & Q\_gist.github.com$\,\to\,$A\_github.com & 51.72\% & 15 \\
 & Q\_npmjs.org$\,\to\,$A\_github.com & 53.85\% & 14 \\
 & Q\_pastebin.com$\,\to\,$A\_github.com & 58.33\% & 14 \\
\npm & Q\_www.npmjs.org$\,\to\,$A\_github.com & 60.00\% & 12 \\
 & Q\_medium.com$\,\to\,$A\_github.com & 66.67\% & 10 \\
 & Q\_yarnpkg.com$\,\to\,$A\_github.com & 50.00\% & 7 \\
 & Q\_nodejs.org, Q\_github.com$\,\to\,$A\_github.com & 75.00\% & 6 \\
 & Q\_www.typescriptlang.org$\,\to\,$A\_github.com & 62.50\% & 5 \\
 & Q\_yeoman.io$\,\to\,$A\_github.com & 62.50\% & 5 \\
 & Q\_registry.npmjs.org, Q\_github.com$\,\to\,$A\_github.com & 62.50\% & 5 \\
 & Q\_ionicframework.com$\,\to\,$A\_github.com & 71.43\% & 5 \\
 & Q\_travis-ci.org$\,\to\,$A\_github.com & 83.33\% & 5 \\ \midrule
 & Q\_i.stack.imgur.com, Q\_docs.microsoft.com$\,\to\,$A\_i.stack.imgur.com & 76.47\% & 13 \\
 & Q\_github.com, Q\_www.nuget.org$\,\to\,$A\_github.com & 50.00\% & 11 \\
\NuGet & Q\_docs.microsoft.com, Q\_stackoverflow.com$\,\to\,$A\_i.stack.imgur.com & 50.00\% & 8 \\
 & Q\_i.stack.imgur.com, Q\_docs.nuget.org$\,\to\,$A\_i.stack.imgur.com & 60.00\% & 6 \\
 & Q\_i.stack.imgur.com, Q\_docs.microsoft.com, Q\_stackoverflow.com$\,\to\,$A\_i.stack.imgur.com & 85.71\% & 6 \\ \bottomrule
\end{tabular}
\end{table*}

\noindent\textbf{Results.} 
Figure~\ref{Fig:RQ6_attribute} shows that external link sharing is the most common practice to resolve PM issues by end-users. This finding is consistent with several previous studies~\cite{gomez2013study, liu2022exploratory}, that also reported  how external links connect to the existing information's and extend the crowd-sourced knowledge. In detail,
\syful{Figure~\ref{Fig:RQ6_attribute}(b) shows that the external links (i.e., {\Maven}: 45.09\% ,  {\npm}: 44.37\%, and {\NuGet}: 42.86\%) are the most frequent attributes in the accepted answers.}
Taking a closer look on the external links as shown in Table~\ref{tab:RQ6_rules}, we observe that end-users share official website links of PMs followed by github and image screenshots as part of information to describe PM issues and accepted answers. 
Moreover, we observed that links referring to the useful resources (i.e., official documentation websites, tutorial etc.) are most frequently shared.
This analysis on both PM questions and their accepted answers reveal the opportunities for tool support and strategies to find the information needed for end-users to resolve their issues while dealing with PM.

\begin{tcolorbox}
    \textbf{Answering RQ$_3$}: 
   To resolve PM issues, external link sharing is the most common practice by end-users. We noticed that links referring to the useful resources (i.e., official documentation websites, tutorial etc.) are most frequently shared.
\end{tcolorbox}

 \section{Implications}
  \label{implications}

The results of this study can help PM communities better focus their efforts on important issues relating to PM tool usage.
 In the following paragraphs, we describe how the results can be used to better guide PM end-users, PM designers, and researchers.

 \textbf{In terms of PM issues and their underlying causes, we find that  most issues are raised by end-users due to a lack of instructions and error messages from PM tools usage.} In light of these findings, it is necessary to better understand what technical background knowledge end-users should have concerning PMs. \syful{ The end-users should look for PM that supports more than just managing packages such as listing available/installed packages, searching, filtering, installing remotely/locally, supporting wildcards, publishing packages easily and learning to deal with packages. Moreover, the target PM should be checked to determine whether it supports customization, including interactive mode that allows the end-user to decide what steps to take during the installation process. The designers of PM are advised to develop a graphical user interface that is intuitive and simple for end-users. For example, when designing a command line interface, end-user commands should be as intuitive as possible. In order to facilitate end-user troubleshooting, the PM designers should provide more working examples along with relevant troubleshooting information.  In addition, they need to make it easy for end-users to locate the information need to resolve problems, for example by providing good error messages. Further research in these areas may also contribute to the improvement of PM tool development process. Our study seeks to identify the challenges faced by end-users of PM tools in order to guide future research.}

 \textbf{Regarding PM issue resolution, we have discovered that sharing external links appears to be the most common practice among Stack Overflow users responding to PM issues.} \syful{We observed that links pointing to useful resources such as official documentations and tutorials are most frequently shared. This finding reveals the opportunities for tool support and strategies to find useful information for addressing PM issues. Researchers could develop approaches that suggest useful information links by automatically analyzing stack traces and logs to help speed up the issue-resolution process. Furthermore, it would be interesting to gain more insights into how PM end-users discover and disseminate knowledge through better understanding of external link sharing practice. We also encourage future research to focus on domain-specific automated question answering by utilizing such crowd-sourced knowledge.}

\section{Threats to Validity}
\label{validity}
This section describes the internal, external, and construct validity threats of our study.

\noindent\textbf{\textit{Internal Validity.}} 
Threats to internal validity relate to experimental bias and error in conducting the analysis.
\syful{We perform manual analysis on random sample since the PM dataset size is large. Our data sampling strategy may pose threat to the study findings. We adopted a random sampling approach since it has been used in many previous publications that also used the same Stack Overflow dataset, including studies on mobile~\cite{rosen2016mobile} and chatbot~\cite{abdellatif2020challenges}. Additionally, an alternative data sampling strategy such as selecting the most popular question posts (those with the highest number of votes), or only focusing on those with answers, may produce more reliable results.
 To mitigate this challenge, we prepare representative samples for three PMs, with a confidence level of 95\% and a interval of 5. Thus, we believe that experimental bias and error in conducting the analysis were reduced.}\\

\noindent\textbf{\textit{External Validity.}} 
Threats to external validity relate to the generalizability of findings. In our study, we focused only on Stack Overflow which is the largest and most popular question-and-answer platform among end-users. The findings of our study may not generalize to other question-and-answer platforms. However, our study is consistent with previous works that also utilized Stack Overflow dataset~\cite{rosen2016mobile, abdellatif2020challenges}.\\

\noindent\textbf{\textit{Construct Validity.}} 
Threats to construct validity are related to potential errors that can occur when extracting data about PMs.
\syful{The first threat is the construct validity of the collected data. We used Stack Overflow tags to identify posts related to the PMs, but some posts may be incorrectly labeled with PM related tags or missing tags. To reduce this threat, we filtered the list of tags by following state-of-the-art approaches~\cite{rosen2016mobile, abdellatif2020challenges}.}

In the qualitative analysis of classifying PM issues and their underlying causes, the questions may be miscoded due to the subjective nature of our coding approach. To mitigate this threat, we took a systematic approach to validate the taxonomy and the comprehension understanding by the three authors in several rounds. Only until the Kappa score reaches 0.87 and 0.96, indicating that the agreement is almost perfect (0.81-1.00), we were able to complete the rest of the sample dataset. \syful{A further risk is the difficulty in obtaining information from PM question and their associated accepted answer posts using regular expressions. With our defined regular expression, we may not be able to filter out all possible information patterns from PM questions and their accepted answers, which could lead to biased results.}

\section{Related Works}
\label{sec:related_work}
Complementary related works are presented throughout this paper. This section describes some additional related research works.

\noindent\textbf{PM Studies.} The prior studies on PM showed that end-users struggled to manage their software dependencies~\cite{bogart2021and, kikas2017structure, decan2019empirical, lungu2010recovering, raemaekers2012measuring, kula2018developers, dietrich2019dependency}. In detail, Bogart et al.~\cite{bogart2021and}, performed multiple case studies on a set of PMs with different tooling policies. They found that end-user practices differ significantly between PMs. Kikas et al.~\cite{kikas2017structure}, analyzed the dependency network of three PMs (i.e., JavaScript, Ruby, and Rust). They reported that there exist significant difference in dependency network structure across language ecosystems. Decan
et al.~\cite{ decan2019empirical} studied several PMs and report that dependency network tend to grow over time in term of size and the number of packages. Lungu et al.~\cite{lungu2010recovering} investigated issues related to dependency graphs and dependency management specifications. They reported that dependencies also exist between projects in a ecosystem. Raemaekers et al.~\cite{raemaekers2012measuring} showed that dependency management involves making cost-benefit decisions related to keeping package dependencies up to date. Kula et al.~\cite{kula2018developers},  mined  end-user responsiveness to existing security awareness mechanisms
on 850K library dependency migrations from 4,659 GitHub projects. They found that end-users were particularly reluctant to update third-party libraries to fix vulnerabilities.  Dietrich et al. ~\cite{dietrich2019dependency} studied seventeen different PMs. 
Their findings reveal that end-users struggle to find a sweet spot between fixed version dependency predictability and flexible dependency agility.

While these studies have shown that end-users struggle to migrate their dependent packages, the common assumption is that PMs broker dependencies without any issues. 
\syful{In this paper, we examine the issues confronted by end-users when using PM tool to manage third-party package dependencies.}

\noindent\textbf{Stack Overflow Studies.} Question-and-Answer platform like Stack Overflow has gained huge research  interest,  with  topics  relating  to  community  dynamics, technical  issues of programmers  and  human  factors~\cite{meldrum2017crowdsourced}.
Several empirical case studies were performed using Stack Overflow data such as improving API documentation and usage scenarios~\cite{venkatesh2016client}, new programming language (Go, Rust, and Swift) related discussion~\cite{chakraborty2021developers}, privacy~\cite{tahaei2020understanding} etc.
Some studies were done on human factors like IT skill~\cite{montandon2021skills}, programmers expertise~\cite{diyanati2020proposed}, etc.
Several tool supports and recommendation models were developed using Stack Overflow data resources such as PostFinder~\cite{rubei2020postfinder}, bug severity prediction model~\cite{tan2020bug} etc. These studies reported that  Stack Overflow data resources are useful to solve developers challenges.

In this paper, the same data source (i.e, Stack Overflow) is used but different from the above mentioned empirical studies. \syful{To the best of our knowledge, there is no prior work that conducted study on PM issues from Stack Overflow. We extracted PM question-and-answer posts from Stack Overflow and perform a series of empirical studies. In this study, we hope to highlight the challenges and information needs associated with using PM for end users, package manager designers, and researchers.}

\section{Conclusion and Future Work}
\label{conclusions}

In this study, we have explored issues faced by end-users when using PMs through an empirical study of content on Stack Overflow question and accepted answer posts.
We carried out a qualitative analysis of 1,131 question and their accepted answer posts from the \Maven, \npm, and \NuGet\ dependency ecosystems to identify issue types, underlying causes, and their resolutions.
We observe that most issues arise from the lack of understanding PM tool usage information rather than specific version updates and compatibility issues.
To resolve PM issues, we observed that sharing external links was the most common practice. Furthermore, we noticed that the most common links shared were those which pointed to useful materials (such as the official documentation websites, tutorials, etc.).
\syful{In conclusion, this study opens up opportunities for future research in the PM area, such as investigating tool support for novice PM end-users, intuitive PM configuration options, and intuitive error messages through stack traces and log files.}

\section*{Acknowledgment}
This work is supported by the Japanese Society for the Promotion of Science (JSPS) KAKENHI Grant Numbers JP20K19774 and JP20H05706.


\begin{thebibliography}{99}
\bibitem{Web:libraries}
libraries.io, “Helping you make more informed decisions about the software you use.” https://libraries\\.io/about, 2020. (Accessed on 05/20/2020).

\bibitem{Web:npmStat}
the npm blog, “npm blog: Next Phase Montage.” https://blog.npmjs.org/post/612764866888007680/\\next-phase-montage, 2020. (Accessed on 05/20/2020).

\bibitem{dietrich2019dependency}
J. Dietrich, D. Pearce, J. Stringer, A. Tahir, and
K. Blincoe, “Dependency versioning in the wild,”
2019 IEEE/ACM 16th International Conference
on Mining Software Repositories (MSR), pp.349–
359, IEEE, 2019.

\bibitem{Cox-ICSE2015}
J. Cox, E. Bouwers, M. Eekelen, and J. Visser,
“Measuring dependency freshness in software sys-
tems,” 2015 IEEE/ACM 37th IEEE Interna-
tional Conference on Software Engineering (ICSE),
pp.109–118, 2015.

\bibitem{bogart2016break}
C. Bogart, C. Kästner, J. Herbsleb, and F. Thung,
“How to break an api: cost negotiation and com-
munity values in three software ecosystems,” Pro-
ceedings of the 2016 24th ACM SIGSOFT Inter-
national Symposium on Foundations of Software
Engineering, pp.109–120, 2016.

\bibitem{kula2018developers}
R.G. Kula, D.M. German, A. Ouni, T. Ishio,
and K. Inoue, “Do developers update their library
dependencies?,” Empirical Software Engineering,
vol.23, no.1, pp.384–417, 2018.

\bibitem{decan2018impact}
 A. Decan, T. Mens, and E. Constantinou, “On
the impact of security vulnerabilities in the npm
package dependency network,” Proceedings of the
15th International Conference on Mining Software
Repositories, pp.181–191, 2018.

\bibitem{decan2019empirical}
A. Decan, T. Mens, and P. Grosjean, “An empirical comparison of dependency network evolutioin seven software packaging ecosystems,” Empirical Software Engineering, vol.24, no.1, pp.381–416,
2019.

\bibitem{islam2021contrasting}
 S. Islam, R.G. Kula, C. Treude, B. Chinthanet,
T. Ishio, and K. Matsumoto, “Contrasting third-
party package management user experience,” 2021
IEEE International Conference on Software Main-
tenance and Evolution (ICSME), pp.664–668,
IEEE, 2021.

\bibitem{treude2011programmers}
C. Treude, O. Barzilay, and M.A. Storey, “How
do programmers ask and answer questions on the
web?(nier track),” Proceedings of the 33rd interna-
tional conference on software engineering, pp.804–
807, 2011.

\bibitem{DBLP:conf/msr/BaltesDT008}
 S. Baltes, L. Dumani, C. Treude, and S. Diehl,
“Sotorrent: reconstructing and analyzing the evo-
lution of stack overflow posts,” Proceedings of
the 15th International Conference on Mining Soft-
ware Repositories, MSR 2018, Gothenburg, Swe-
den, May 28-29, 2018, ed. A. Zaidman, Y. Kamei,
and E. Hill, pp.319–330, ACM, 2018.

\bibitem{abdellatif2020challenges}
 A. Abdellatif, D. Costa, K. Badran, R. Abdalka-
reem, and E. Shihab, “Challenges in chatbot de-
velopment: A study of stack overflow posts,” Pro-
ceedings of the 17th International Conference on
Mining Software Repositories, pp.174–185, 2020.

\bibitem{rosen2016mobile}
C. Rosen and E. Shihab, “What are mobile devel-
opers asking about? a large scale study using stack
overflow,” Empirical Software Engineering, vol.21,
no.3, pp.1192–1223, 2016.

\bibitem{viera2005understanding}
 A.J. Viera, J.M. Garrett, et al., “Understand-
ing interobserver agreement: the kappa statistic,”
Fam med, vol.37, no.5, pp.360–363, 2005.

\bibitem{HataICSE2019}
 H. Hata, C. Treude, R.G. Kula, and T. Ishio, “9.6
million links in source code comments: Purpose,
evolution, and decay,” Proceedings of the 41st In-
ternational Conference on Software Engineering,
p.1211–1221, 2019.

\bibitem{nagaria2020software}
 B. Nagaria and T. Hall, “How software developers
mitigate their errors when developing code,” IEEE
Transactions on Software Engineering, 2020.

\bibitem{mangul2019challenges}
 S. Mangul, T. Mosqueiro, R.J. Abdill, D. Duong,
K. Mitchell, V. Sarwal, B. Hill, J. Brito, R.J.
Littman, B. Statz, et al., “Challenges and rec-
ommendations to improve the installability and
archival stability of omics computational tools,”
PLoS biology, vol.17, no.6, p.e3000333, 2019.

\bibitem{argelich2008cnf}
 J. Argelich and I. Lynce, “Cnf instances from the
software package installation problem.,” RCRA,
2008.

\bibitem{decan2017empirical}
 A. Decan, T. Mens, and M. Claes, “An empiri-
cal comparison of dependency issues in oss pack-
aging ecosystems,” 2017 IEEE 24th International
Conference on Software Analysis, Evolution and
Reengineering (SANER), pp.2–12, IEEE, 2017.

\bibitem{abate2015mining}
 P. Abate, R. Di Cosmo, L. Gesbert, F. Le Fes-
sant, R. Treinen, and S. Zacchiroli, “Mining com-
ponent repositories for installability issues,” 2015
IEEE/ACM 12th Working Conference on Mining
Software Repositories, pp.24–33, IEEE, 2015.

\bibitem{gomez2013study}
 C. Gómez, B. Cleary, and L. Singer, “A study of
innovation diffusion through link sharing on stack
overflow,” 2013 10th Working Conference on Min-
ing Software Repositories (MSR), pp.81–84, IEEE,
2013.

\bibitem{liu2022exploratory}
J. Liu, H. Zhang, X. Xia, D. Lo, Y. Zou, A.E.
Hassan, and S. Li, “An exploratory study on
the repeatedly shared external links on stack
overflow,” Empirical Software Engineering, vol.27,
no.1, pp.1–32, 2022.

\bibitem{bogart2021and}
 C. Bogart, C. Kästner, J. Herbsleb, and F. Thung,
“When and how to make breaking changes: Poli-
cies and practices in 18 open source software
ecosystems,” ACM Transactions on Software Engi-
neering and Methodology (TOSEM), vol.30, no.4,
pp.1–56, 2021.

\bibitem{kikas2017structure}
 R. Kikas, G. Gousios, M. Dumas, and D. Pfahl,
“Structure and evolution of package depen-
dency networks,” 2017 IEEE/ACM 14th Interna-
tional Conference on Mining Software Repositories
(MSR), pp.102–112, IEEE, 2017.

\bibitem{lungu2010recovering}
 M. Lungu, R. Robbes, and M. Lanza, “Recover-
ing inter-project dependencies in software ecosys-
tems,” Proceedings of the IEEE/ACM interna-
tional conference on Automated software engineer-
ing, pp.309–312, 2010.

\bibitem{raemaekers2012measuring}
 S. Raemaekers, A. Van Deursen, and J. Visser, “Measuring software library stability through historical version analysis,” 2012 28th IEEE In ternational Conference on Software Maintenance (ICSM), pp.378–387, IEEE, 2012.

\bibitem{meldrum2017crowdsourced}
S. Meldrum, S.A. Licorish, and B.T.R. Savarimuthu, “Crowdsourced knowledge on stack overflow: A systematic mapping study,” Proceedings of the 21st International Conference on Evaluation and Assessment in Software Engineering, pp.180–185, 2017.

\bibitem{venkatesh2016client}
P.K. Venkatesh, S. Wang, F. Zhang, Y. Zou, and
A.E. Hassan, “What do client developers concern
when using web apis? an empirical study on de-
veloper forums and stack overflow,” 2016 IEEE In-
ternational Conference on Web Services (ICWS),
pp.131–138, IEEE, 2016.

\bibitem{chakraborty2021developers}
 P. Chakraborty, R. Shahriyar, A. Iqbal, and
G. Uddin, “How do developers discuss and support
new programming languages in technical q\&a site?
an empirical study of go, swift, and rust in stack
overflow,” Information and Software Technology,
vol.137, p.106603, 2021.

\bibitem{tahaei2020understanding}
M. Tahaei, K. Vaniea, and N. Saphra, “Under-
standing privacy-related questions on stack over-
flow,” Proceedings of the 2020 CHI Conference on
Human Factors in Computing Systems, pp.1–14,
2020.

\bibitem{montandon2021skills}
J.E. Montandon, C. Politowski, L.L. Silva, M.T.
Valente, F. Petrillo, and Y.G. Guéhéneuc, “What
skills do it companies look for in new developers?
a study with stack overflow jobs,” Information and
Software Technology, vol.129, p.106429, 2021.

\bibitem{diyanati2020proposed}
A. Diyanati, B.S. Sheykhahmadloo, S.M. Fakhrah-
mad, M.H. Sadredini, and M.H. Diyanati, “A pro-
posed approach to determining expertise level of
stackoverflow programmers based on mining of
user comments,” Journal of Computer Languages,
vol.61, p.101000, 2020.

\bibitem{rubei2020postfinder}
R. Rubei, C. Di Sipio, P.T. Nguyen, J. Di Rocco,
and D. Di Ruscio, “Postfinder: Mining stack
overflow posts to support software developers,”
Information and Software Technology, vol.127,
p.106367, 2020.

\bibitem{tan2020bug}
Y. Tan, S. Xu, Z. Wang, T. Zhang, Z. Xu, and
X. Luo, “Bug severity prediction using question-
and-answer pairs from stack overflow,” Journal of
Systems and Software, vol.165, p.110567, 2020.

\end{thebibliography}
\bibliographystyle{ieicetr}

\profile[Authors_Photo/Syful.JPG]{Syful Islam}{received the M.E., and PhD degree in Engineering from Nara Institute of Science and Technology, Japan.  At present, he is working as an assistant professor at Noakhali Science and Technology University, Bangladesh. His research interests include software ecosystem, mining Stack Overflow, etc.}

\profile[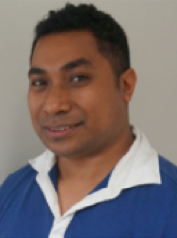]{Raula Gaikovina Kula}{is currently an assistant professor at Nara Institute of Science and technology. In 2013, he graduated with a PhD. from Nara Institute of Science and Technology, Japan. He is currently an active member of the IEEE Computer Society and ACM. His research interests include repository mining, code review, software libraries and visualizations.}

\profile[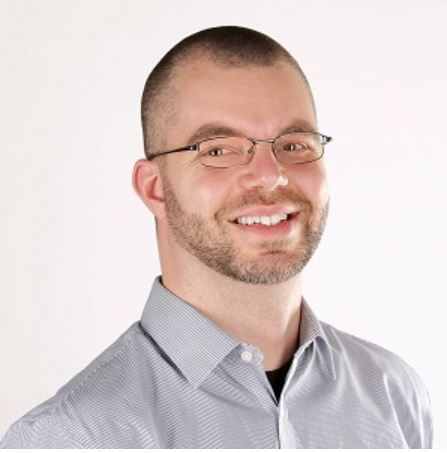]{Christoph Treude }{is a Senior Lecturer in Software Engineering in the School of Computing and Information Systems at the University of Melbourne. His research combines empirical studies with the innovation of tools and approaches that take the wide variety of natural language artefacts in software repositories into account.}

\profile[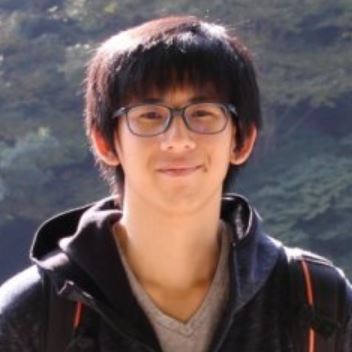]{Bodin Chinthanet }{is currently a Specially Appointed Assistant Professor in Software Engineering Laboratory under the supervision of Professor Kenichi Matsumoto, Nara Institute of Science and Technology (NAIST). His research interests include empirical software engineering and mining software repositories.  }

\profile[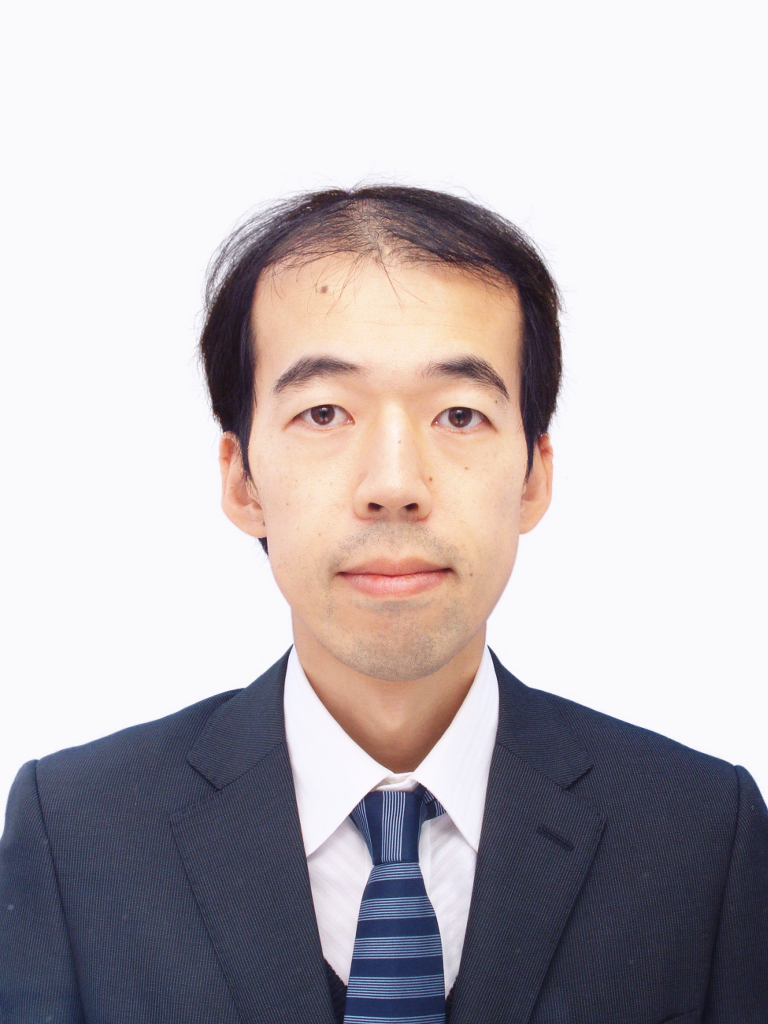]{Takashi Ishio}{received the Ph.D degree in information science and technology from Osaka University in 2006.
He was a JSPS Research Fellow from 2006-2007.
He was an assistant professor at Osaka University from 2007-2017.
He is now an associate professor of Nara Institute of Science and Technology.
His research interests include program analysis, program comprehension, and software reuse.
He is a member of the IEEE, ACM, IPSJ and JSSST.}

\profile[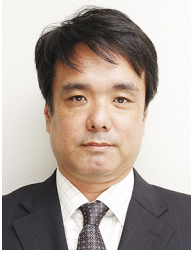]{Kenichi Matsumoto}{received the B.E., M.E., and PhD degrees in Engineering from Osaka University, Japan, in 1985, 1987, 1990, respectively. Dr. Matsumoto is currently a professor in the Graduate School of Information Science at Nara Institute Science and Technology, Japan. His research interests include software measurement and software process. He is a senior member of the IEEE and a member of the IPSJ and SPM.}

\end{document}